\documentclass[journal]{IEEEtran}

\usepackage{amsmath,amssymb,amsfonts}
\usepackage{tabularx}
\usepackage[utf8]{inputenc} % allow utf-8 input
\usepackage[T1]{fontenc}    % use 8-bit T1 fonts
\usepackage{url}            % simple URL typesetting
\usepackage{booktabs}       % professional-quality tables
\usepackage{amsfonts}       % blackboard math symbols
\usepackage{nicefrac}       % compact symbols for 1/2, etc.
\usepackage{microtype}      % microtypography
\usepackage{graphicx}
\usepackage{float}
\restylefloat{table}
\usepackage{multicol}
\usepackage{caption}
\usepackage{subcaption}
\usepackage{amsmath}
\usepackage{algorithm}
\usepackage{algpseudocode}
\usepackage{tikz}
\usetikzlibrary{trees}
\usepackage{listings}

\DeclareMathOperator*{\argmax}{arg\,max}  % in your preamble
\usepackage{textcomp}

\usepackage{xcolor,soul,framed} %,caption

\usepackage[noadjust]{cite}
\begin{document}
\bstctlcite{IEEEexample:BSTcontrol}
    \title{test}
\title{Multiagent Reinforcement Learning based Energy Beamforming Control}
\author{ Zongqiang Pang,Liping Bai ~\IEEEmembership{Member,~IEEE,} \thanks{Nanjing Unversity of Posts and Telecommunications, College of Automation \& College of Artificial Intelligence, Nanjing, Jiangsu,210000 China email:zqpang@njupt.edu.cn}}

% ====================================================================
\maketitle

% === ABSTRACT ====================================================================
% =================================================================================
\begin{abstract}
%\boldmath
Ultra low power devices make far-field wireless power transfer a viable option for energy delivery despite the exponential attenuation. Electromagnetic beams are constructed from the stations such
that wireless energy is directionally concentrated around the ultra low power devices. Energy beamforming faces different challenges compare to information beamforming due to the lack of feedback on channel state. Various methods have been proposed such as one-bit channel feedback to enhance energy beamforming capacity, yet it still has considerable computation overhead and need to be computed centrally. Valuable resources and time is wasted on transfering control information back and forth. In this paper, we propose a novel multiagent reinforcement learning(MARL) formulation for codebook based beamforming control. It takes advantage of the inherienntly distributed structure in a wirelessly powered network and lay the ground work for fully locally computed beam control algorithms. Source code can be found at \url{https://github.com/BaiLiping/WirelessPowerTransfer}.
\end{abstract}

% === KEYWORDS ====================================================================
% =================================================================================
\begin{IEEEkeywords}
Multiagent Reinforcement Learning,MARL, Wireless Power Transfer, Beamforming
\end{IEEEkeywords}

% \fi
\IEEEpeerreviewmaketitle

% ====================================================================
% ====================================================================
% ====================================================================

% === I. INTRODUCTION =============================================================
% =================================================================================
\section{Introduction}
\IEEEPARstart{W}{ireless} power transfer(WPT) can be divided into near-field WPT with inductive coupling, magnetic resonant coupling or capacitive coupling {8357386} and far-field WPT with electromagnetic power beams. \cite{8246215} Compare to near-field WPT, the far-field option has considerable attenuation,  yet the increased application of ultra low power devices such as RFID, low power sensor networks \cite{7578025}, together with various forms of joint wireless information and power transfer technology such as Simultaneous Wireless Information and Power Transfer (SWIPT) \cite{8214104}, Wirelessly Powered Communication Networks (WPCNs) \cite{7462480}, Wirelessly Powered Backscatter Communication (WPBC) \cite{7842391} has made far-field WPT an important tool for powering those devices. Current far-field WPT technology can effectively transfer tens of microwatts of RF power to wireless devices from a distance of more than 10 meters. \cite{7462480}

For far-field WPT to be as effective as it can be, directional RF phased array, a group of radiating elements whose phase and magnitude can be controlled to generate a directional beam pattern, \cite{665} is utilized to increase the directional gain of power transfer. Digital phased control has high fidelity and is mostly used for communication systems such as 5G Antenna. However, its energy and thermal cost make it prohibitively expensive for other applications. Analog phased control utilizes RF chain to systematically shift the phase discretely. In this paper, we focus on analog phased array.

There are two kinds of control algorithms for analog beamformer, one is adaptive beamforming, which can adjust according to various channel conditions, but it is expensive in terms of data collection and computational time. A less versatile control algorithm is switch-based control. There are set of predetermined codes for beamforming control. An exhaustive search is performed to find the "optimal code" for the given circumstances\cite{5262295} \cite{1237152}. The codebook exhaustive search algorithm or codebook based beam training process still has a large overhead, particularly for a multi-station scenario. In previous works reinforcement learning based solutions have been proposed where the multi-armed bandit framework \cite{8662770} or Q-learning framework \cite{Cui2019SecureWC} was used to render the process more effective.

In this paper, beamforming control is formulated as a multi-agent reinforcement learning problem. Rollout algorithm proposed by Dimitri P. Bertsekas \cite{Bertsekas2019MultiagentRA} is utilized to properly trade-off action space complexity and state-space complexity, hence reducing the learning time. This paper is arranged as the following. In section II, the system model described. In section III, the setup of multiagent reinforcement learning is introduced. In section IV, the problem of WPT is seen through the lense of multiagent reinforcement learning and the simulation result is presented.

% === II. Harmonically-Terminated Power Rectifier Analysis ========================
% =================================================================================
\section{Enegy Beamforming}
\subsection{Uniform Linear Array}
\begin{figure}[H]
\centering
\includegraphics[width=0.3\textwidth]{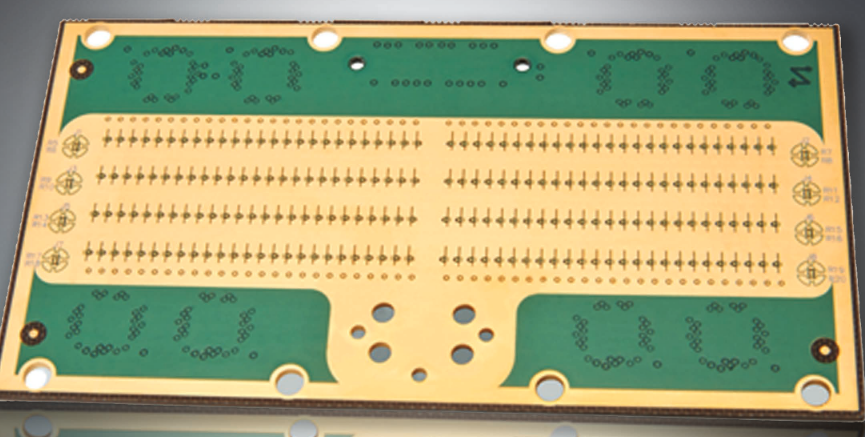}
\caption{Pivotal 39GHz Beamformer}
\label{fig:pivotal}
\end{figure}

The theories of phased array were fully formulated during the WWII era where an array of radars was deployed to detect an accurate angle of arrival \cite{5237174}. Today, phased array hardware is widely available as a commercial product as shown in Figure \ref{fig:pivotal}. Together with various forms of Space Time Signal Processing(STSP), phased array and the beamforming technology has become the enabling components for future communication networks. There are different configurations of arrays, in this paper, we only consider one dimension uniform linear array.

\subsection{Channel Model}

Suppose there are p propogation path from transmitter to receiver. the gain for each path is denoted by $\alpha_i$. The channel is modeled as a sum of each path. When Line of Sight(LoS) not available and the number of path is large, Rayleigh fading and a pleathera of channel modelling techniques can be applied to capture Non Line of Sight channel gain. In this paper, we only consider the environment with direct Line of Sight transmission and no reflection path.

\begin{equation}
h=\sum_{i=1}^p \alpha_i e^{-j2\pi \frac{d_i}{\lambda}}
\end{equation}

\begin{figure}[H]
\centering
\includegraphics[width=0.2\textwidth]{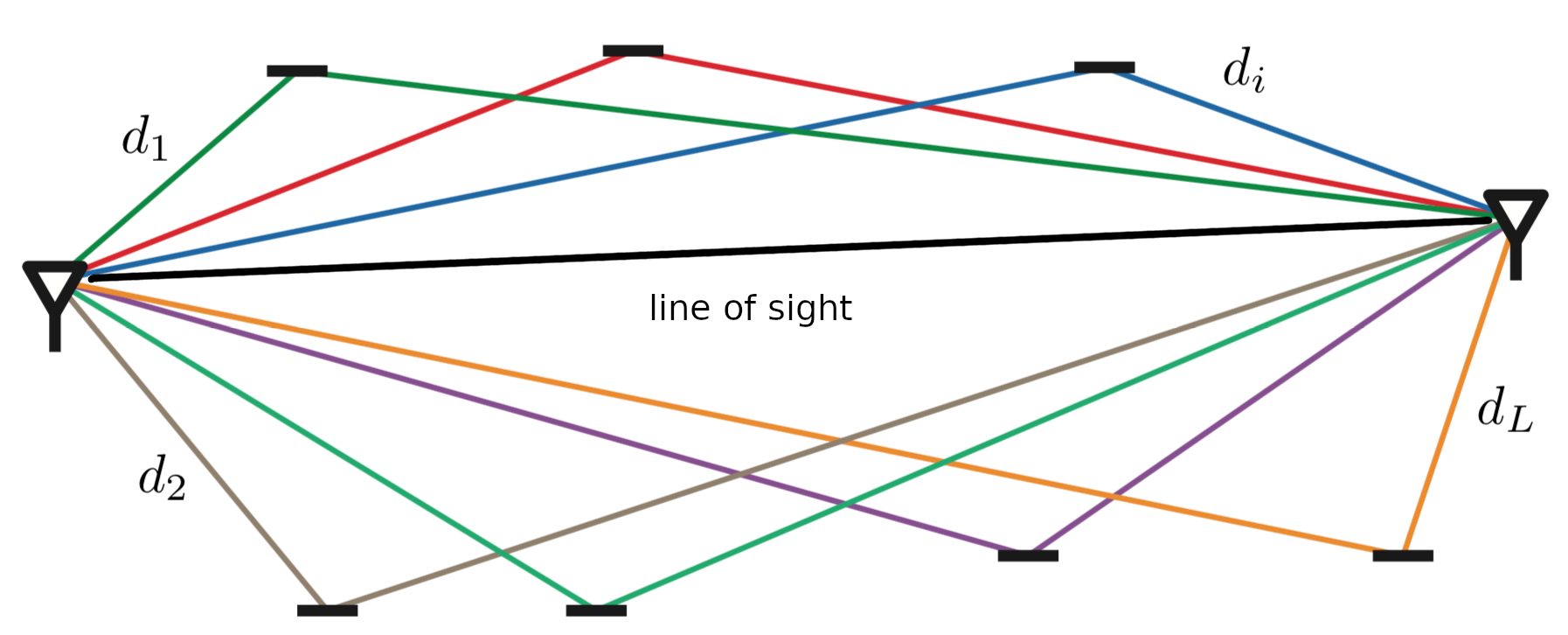}
\caption{Multipath Channel Model}
\label{fig:channel}
\end{figure}

\subsection{beamforming codebook $\mathcal{F}$}

Let Angle of Departure(AoD) be denoted as $\varphi$. Suppose the the range of adjustment for the beamformer is $\zeta$ degrees and is deicretized into N portions, each with the angle adjustment of $\frac{\zeta}{N}$ degree. For the $i^{th}$ code in the codebook with AoD of $\varphi_i$, the beamforming vector is computed as the following:
\begin{equation}
\textbf{f} := \textbf{a}(\varphi_i)= [1,e^{jdcos(\varphi_i)},...,e^{jd(M-1)cos(\varphi_i)}]^T
\end{equation}

\subsection{System Model}

\begin{figure}[H]
    \centering
    \includegraphics[width=0.3\textwidth]{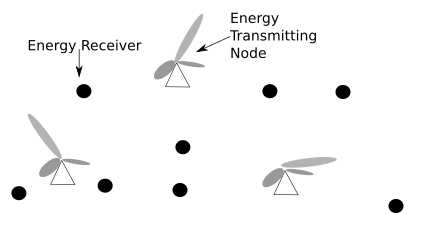}
    \caption{Wirelessly Powered Network}
    \label{fig:MIMO}
\end{figure}

The schematics of wirelessly powered communication network is shown in Figure \ref{fig:MIMO}. L energy transmitting node, each equipped with M radiating elements arranged as a uniform linear array, transmit power to K energy receivers scattered in an open field.

One challenge for energy beamforming is lack of channel information. Time Division Duplex is a common strategy to implement joint communication and energy transfer, where the wireless power transfer happens from time 0 to P and wireless information transfer(WIT) happens from time P to T. Therefore, no information can be sent before enough energy is stored during the power transfer phase. Ideally, pilots signals should be sent and decoded systematically for channel estimation. \cite{1597555}, yet this function is not available for energy beamforming because WIT and WPT functions are realized with separate circuits \cite{7462480}, where the latter does not provide decoding capacity. Previous works have proposed methods of channel estimation with only one-bit feedback \cite{6884811}, we would adopt this minimum feedback scheme in this paper.

Because the energy beamforming signal \textbf{x} does not carry any information, it is assumed to be independent sequences with zero mean and unit variance. \cite{6884811} Furthermore, because we consider the beamformer to be analog, $x_1$ to $x_M$ are the same signal $x \in \mathbb{C}$.The power in noise is significantly weaker than the energy signal, therefore it can be ignored for practical purposes.

Let $y_{j,p}$ denote the signal received on the $j^{th}$ receiver from the $p^{th}$ transmitter. $x_p$ be the signal transmitted from node p. $\textbf{f}_p$ be the beamforming code for node p.  $h_{i,p}^j$ denote the line of signt channel connecting $j^{th}$ receiver to $i^{th}$ radiating element from $p^{th}$ transmitting node.

\[
y_{j,p}=
\begin{bmatrix}
h_{1,p}^j & h_{2,p}^j & ... & h_{M,p}^j\\
\end{bmatrix}
\begin{bmatrix}
f_{1,p}\\
.\\
.\\
f_{M,p}\\
\end{bmatrix}
x_p
\]

In this paper, we assume that the radiating elements from the same node share the same path gain $\alpha$. Let $\alpha_{j,p}$ denote path gain for line of sight channel connecting $j^{th}$ receiver to $p^{th}$ transmitter.$\varphi_{j,p}$ denote the Angle of Departure connecting $j^{th}$ receiver to $p^{th}$ transmitter. $\textbf{a}(\varphi_{j,p})^*=[1,e^{jdcos(\varphi_{j,p}},...,e^{jd(M-1)cos(\varphi_{j,p})})]$. Therefore:
\begin{equation}
y_{j,p}=\alpha_{j,p} \textbf{a}(\varphi_{j,p})^* \textbf{f}_p x_p
\end{equation}
The received signal on the $j^{th}$ receiver is a summation of signals delievered from all the transmitters to this receiver.
\begin{equation}
y_j=\sum_{p=1}^L \alpha_{j,p} \textbf{a}(\varphi_{j,p})^* \textbf{f}_p x_p
\end{equation}
Let the wireless power transfer happening for duration P. Received energe on the $j^{th}$ receiver for duration P is:
\begin{equation}
e_j=\int_0^P |y_j(t)|^2 dt=\int_0^P |\sum_{p=1}^L \alpha_{j,p} \textbf{a}(\varphi_{j,p})^* \textbf{f}_p x_p(t)|^2 dt
\end{equation}

\subsection{Wireless Power Transfer}

The objective of energy beamforming control is to choose a beamforming code for each energy transmitting node such that the total received power is maximized while satisfying the minimum energy requirement of each energy receiver.

\begin{equation*}
    \begin{aligned}
        & \underset{ \textbf{f}_p, \forall p}{\text{maximize}}
        && \displaystyle\sum_{j \in \{1,2,...,L\}} e_j\\
        & \text{subject to}
        && \textbf{f}_p \in \mathcal{F}, \forall p;\\
        &&& e_j \geq e_{min}\\
    \end{aligned}
\end{equation*}

\section{Reinforcement Learning}

\subsection{problem setup}
The impetus of reinforcement learning is that an agent can learn by interacting with the environment. In the intersection between control, optimization, and learning, the problem have different mathematical formulations. Here, we follow the problem setup proposed by Richard Sutton in his book Introduction to Reinforcement Learning. \cite{10.5555/551283}

Agent can observe the state at each step, denoted as $ S_{t} $, where t is the $t^{th}$ step taken. For our discussion, we focus only on the subset of problems where state s is fully observable by the agent. There are action choices for each state denoted as $ A_{t} $. A reward is given for each action taken at step t denoted as $ R_{t} $. The terminal step is denoted as t=T. For an episodic problem, T is a finite number, for a non-episodic problem, T=$\infty$

An episode of data is registered as an alternating sequence of state, action, and reward:

$$ S_{0}, A_{0}, R_{0}, S_{1}, A_{1}, R_{1}.......S_{T-1},A_{T-1},R_{T-1},S_{T},A_{T},R_{T} $$

Gain at step t is defined as the accumulative reward the agent can get from step t onward. A discounting factor $\gamma$ between 0 to 1 is introduced to incorporate the sense of time, much like how interest rate encodes time in financial systems:

\begin{equation}
G_{t} := R_{t}+\gamma R_{t+1}+\gamma ^2 R_{t+2}+...+\gamma^{T-t}R_{T}
\end{equation}

This can be written in its recursive form, known as Bellman Equation, which is the basis for an iteratively implemented backward induction algorithm:

\begin{equation}
        G_{t}=R_{t}+\gamma G_{t+1}
    \label{bellman}
\end{equation}
Transition matrix is intruduced to encode the stochastidy in the environmental dynamics. Transaction Matrix $\mathcal{P}$ is defined as:

\begin{equation}
    \mathcal{P}_{ss'}^a := Pr\{S_{t+1}=s'|S_{t}=s,A_{t}=a\}
\end{equation}

State/Action Function q(s,a) is definied as expected gain starting from state s by taking action a:

\begin{align}
\begin{split}
    q(s,a) :&= \mathbb{E}\{G_t|S_t=s,A_t=a\}\\
    &=\mathbb{E} \{\sum_{k=0}^{T-t} \gamma ^k R_{t+k+1}| S_t=s,A_t=a\}\\
\end{split}
\end{align}
Policy is defined as:
\begin{equation}
    \pi(s,a):=Pr(A=a|S=s)
\end{equation}
Optimal Policy is defined as:
\begin{equation}
    \pi^*(s):=\argmax_a q(s,a)
\end{equation}

Value Function v(s) is defined as the expected gain starting from state s:

\begin{align}
\begin{split}
    v(s) :&=  \mathbb{E}\{G_t|S_t=s\}\\
    &=\mathbb{E} \{\sum_{k=0}^{T-t} \gamma ^k R_{t+k+1}| S_t=s\}\\
    &=\sum_{a \in \mathcal{A}} \pi(s,a) q(s,a)\\
\end{split}
\end{align}

\subsection{Without Approximation}
One obvious approach to learning is to statistically construct a model of the environment, which is called Model-Based Learning. The most primitive form of model-based learning is Bellman Equation based backward induction. Statistical tactics, such as maximum likelihood, Bayesian methods, etc., can be deployed to approximate the model with the least amount of sampling. However, since the environment is implicitly embedded in v(s) and q(s,a), the model building process can be circumvented entirely, hence Model-Free Learning. Depending on whether the iteration rules is policy dependent, model-free learning can be subdivided into on-policy learning and off-policy learning.

One hindrance to the implementation of the brute force backward induction is its memory requirement. A more effective approach is to update q value and v value after one episode, one step, or n steps. They are called Monte Carlo Method, Temporal Difference Method, and $\lambda(n)$ Method respectively.

For online learning, $\epsilon$-greedy Policy $\pi_{\epsilon}(s)$ is frequently deployed to balance exploration and exploitation, such that the environment can be encoded most efficiently. $\epsilon$ is initiated set to 1 and then asymptotically goes to 0 as the episode counts increases.

\begin{equation*}
    \pi_{\epsilon}(s,a) = \begin{cases}
        1-\epsilon+\frac{\epsilon}{|A|}& \displaystyle\argmax_{a} q_{\epsilon}(s,a)\\
        \frac{\epsilon}{|A|}& \text{otherwise}\\
           \end{cases}
\end{equation*}

\subsection{With Approximation}
When the problem gets complex, state S becomes a rather large vector and function approximation with neuro networks can be utilized to facilitate learning. Reinforcement learning as a self-sustaining mathematical framework has been refined by Rich Sutton et al. since the 1980s. Only recently, the progress made with Deep Learning has been applied to the realm of Reinforcement Learning \cite{Mnih2013PlayingAW}, rendering the computation tenable with existing hardare.

Let the value function and state/action function be parameterized with $\textbf{w}:  \hat{v}(s,\textbf{w}) \approx v(s)$ and $\hat{q}(s,a,\textbf{w}) \approx q(s,a) $

Let the $i^{th}$ iteration of parameter be denoted as $\textbf{$w_i$}$. The Loss Function $\mathcal{L}(\textbf{$w_i$})$ is defined as the following:
\begin{equation}
    \mathcal{L}(\textbf{$w_i$}) := \mathbb{E}\{[v(s)-\hat{v}(s,\textbf{$w_i$})]^2\}
    \label{v_loss}
\end{equation}

\begin{equation}
    \mathcal{L}(\textbf{$w_i$}) := \mathbb{E}\{[q(s,a)-\hat{q}(s,a,\textbf{$w_i$})]^2\}
    \label{q_loss}
\end{equation}
While the real value of v(s) and q(s,a) are not knowable, it can be approximated:
\begin{equation}
    v(s) \approx \sum_{a \in A} R(s,a)+\gamma v(s',\textbf{w})
    \label{v_approx}
\end{equation}
\begin{equation}
    q(s,a) \approx r+\gamma \argmax_{a} q(s',a',\textbf{w})
    \label{q_approx}
\end{equation}

The Gradient of weighing paramter \textbf{w} can be derived from \ref{v_loss} and \ref{q_loss} with the real values substituted by \ref{v_approx} and \ref{q_approx} respectively. By convention, constant is omitted. Parameter is updated following Gradient Descent:
\begin{equation}
\textbf{w}_{i}=\textbf{w}_{i-1}-\nabla_{w_{i-1}} \mathcal{L}(w_{i-1})
\end{equation}
\subsection{Policy Gradient Methods}
Policy $\pi(s)$ can be written as a function parameterized by $\theta$ with s as input and a smooth distribution overall all actions as output.By adjusting parameter $\theta$ we can adjust the distribution over action choices for different states. This style of learning is called policy gradient-based learning.

Let us register a path sequence taken by the agent as $\tau$ such that the sequence is denoted as \{$S_{\tau 0},A_{\tau 0}, R_{\tau 0}...S_{\tau T},A_{\tau T},R_{\tau T}$\}. the gain of sequence $\tau$ is defined as the gain of this entire sequence of state, action, reward:
\begin{equation}
    G(\tau):=\displaystyle\sum_{t=0}^{T}\gamma^t R_t
\end{equation}

Denote $P(\tau,\theta)$ as the probability that path $\tau$ is travesed when the policy is parameterized by $\theta$. The Objective Function can be defined in various ways. Here we adopt the definition as the following:
\begin{equation}
    U(\theta)=\sum_{\tau}P(\tau,\theta)G(\tau)
\end{equation}

The objective of the policy gradient method is to find the parameter $\theta$ to maximize the objective function.

The gradient of aforementioned utility function is:
\begin{equation}
    \nabla_{\theta} U(\theta)= \nabla_{\theta}\sum_{\tau}P(\tau,\theta) G(\tau)
\end{equation}

A mathematical sleight of hand called Importance Sampling is deployed to convert this theoretical expression of gradient into something that is algorithmically feasible.

\begin{align}
\begin{split}
\nabla_{\theta} U(\theta) \approx \frac{1}{N}\displaystyle\sum_{\tau=1}^{N}\displaystyle\sum_{t=0}^{T-1} \nabla_{\theta}ln\pi_{\theta}(s,a)|_{\theta_{old}}[q^{\pi_{\theta_{old}}}(s,a)-b]\\
\end{split}
\end{align}

We can use stochastic gradient descent(SGD) method to update $\theta$:
\begin{equation}
    \theta=\theta_{old}-\alpha \nabla_{\theta}ln\pi_{\theta}(s,a)|_{\theta_{old}}[q^{\pi_{\theta_{old}}}(s,a)-b]
\end{equation}

Actor-Critic Method takes advantage of both policy gradient and function approximation to build a bootstrap structure that lead up to fast convergence. state/action function for policy $\pi_{\theta}(s)$ is approximated by $q^{\pi_{\theta}}(s,a,\textbf{w})$.Baseline b is introduced into the bootstrap stracture to foster convergence. Different algorithms define baseline differently. In advantage Actor-Critic algorithm, baseline is defined as a value function based on $\pi_{\theta}$.Because the SGD updating process does not rely on the ordering of things, it is obvious that some of the aforementioned computations can be done asynchronously. Asynchronous Advantage Actor-Critic (A3C) is proven one of the most effective agents for renforcement learning, and is the one we will use in this paper.

\subsection{Multiagent Reinforcement Learning}
$A_t=\{A_t^1,A_t^2,...,A_t^M\}$ M is the number of agents. The action space is cartician product of action choices available to each agent. $A_t(s)=A_t^1(s) \times A_t^2(s) \times ... \times A_t^M(s)$, which grows exponentially as the number of agents grows.

The Rollout method proposed by Dimitri Bertsekas breakdown this collective decision into its sequential components, reducing the complexity of action space while increasing the complexity of state space. It is proven that the intermediate state rollout method yields the same result as does the regular method. \cite{Bertsekas2019MultiagentRA}

\begin{figure}[H]
\centering
\includegraphics[width=0.3\textwidth]{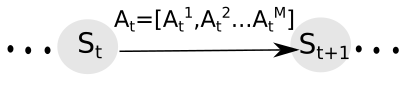}
\caption{Without Rollout}
\label{fig:without}
\end{figure}

Without intermidiate state rollout, the sequences of data collected is:
...$S_t, A_t, R_t, S_{t+1}$... as shown in Figure \ref{fig:without}

\begin{figure}[H]
\centering
\includegraphics[width=0.55\textwidth]{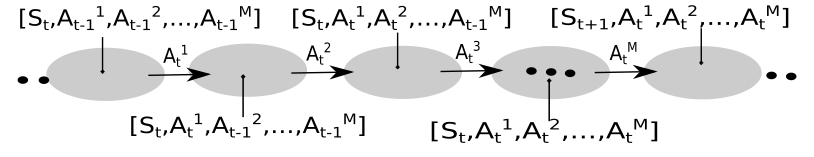}
\caption{With Rollout}
\label{fig:with}
\end{figure}

The intermediate states rollout technique converting action space complexity into state-space complexity by introducing intermediate states, denoted as $S_t^k$ where k goes from 1 to M-1. The sequence of data is now: ...$S_t,A_t^1,R_t^1,S_t^1,A_t^2,R_t^2, S_t^2, ... , S_t^{M-1},A_t^M,R_t^M,S_{t+1}$... as shown in Figure \ref{fig:with}

where $S_t^k=(S_t^{k-1},A_t^k)$

suppose each agent has N choices. This formulation reduces the size action space from $N^M$ to $N \times M$.

\section{Beamforming as a Multiagent Reinforcement Learning problem}
\subsection{Environment}
\begin{figure}[H]
\centering
\includegraphics[width=0.3\textwidth]{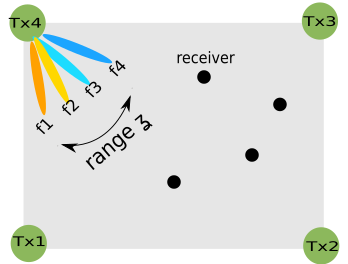}
\caption{Environment}
\label{fig:environment}
\end{figure}

The wirelessly powered communication network has L energy transmitting stations positioned at the corner of a 30m x 30m field. K energy receivers randomly scattered between 1m to 29m as illustrated by Figure \ref{fig:environment}. 0.5s of energy transfer is followed with 0.5s of information transfer. Assume no energy leftover at each cycle, such that at the beginning of the next energy transfer interval, the remaining power at each energy receiver is 0.

\begin{small}
\begin{center}
\begin{tabular}{ c c }
\hline
Number of Energy Transmitting Nodes & L=4\\
\hline
&$TX_1$(0,0)\\
&$TX_2$(30,0)\\
{Positions of Trasmitting Node}&$TX_3$(30,30)\\
&$TX_4$(0,30)\\
\hline
Number of Radiation Elements per Trasmitting Node & M=64\\
\hline
Energy Carrier Frequency & 8M Hz\\
\hline
Field of Energy Receivers & 30m x 30m\\
\hline
Number of Energy Receivers & K \\
\hline
Energy Transfer Time & 0.5s \\
\hline
Information Transfer Time & 0.5s\\
\hline
Maximum Number of Steps & 100\\
\hline

\end{tabular}
\end{center}
\end{small}

Observation Space:\{$e_1,e_2,...e_K,c_1,c_2,c_3,c_4$\}
where $e_j$ is the energy received at the $j^{th}$ receiver, $c_i$ is the codebook choice for the $i^{th}$ energy emitting node.

Reward: If $e_j$ < $e_min$, reward is deducted by 50 points each. If $e_{total}^new$ > $e_{total}^new$, reward is increased by 100 points. If $e_{total}^new$ < $e_{total}^new$, reward is deduced by 300 points.

\subsection{A3C agent}
\begin{small}
\begin{center}
\begin{tabular}{ c c }
\hline
Layers of Actor Network & 3\\
\hline
Layers of Critic Network & 3\\
\hline
Learning Rate for Actor & $\alpha_a$=0.1\\
\hline
Learning Rate for Critic & $\alpha_c$=0.1\\
\hline
Discount Rate & $\gamma$=0.9 \\
\hline
Action Function & Softmax \\
\hline

\end{tabular}
\end{center}
\end{small}

\subsection{Simulation Result}
\section{Conclusion}
In this paper, we demonstrated the possibility to formulate WPT as a multiagent reinforcement learning problem, this lays the groundwork for further study towards fully locally computed control algorithms for wirelessly powered communication networks. Instead of group actions of all agents together, a multiagent rollout approach sees things sequentially, action taken by one agent becomes part of the state of another. This framework deduces the dimension of action space from exponential growth to multiplicative growth, and it can be applied to other problems. The most recent incarceration of beamforming technology is a passive reflective surface, or Intelligent Reflective Surface(IRS), where the reflective components are in the thousands. The multiagent approach proposed in this paper could be applied to IRS control as well, which should be a fruitful topic of future studies.
\bibliographystyle{IEEEtran}
\bibliography{Bibliography}
%\printbibliography
\end{document}